\documentclass[11pt]{article}
\usepackage{lscape,epsfig,amsmath,amssymb,latexsym,
	psfig,psfrag,graphicx,feynarts}
\usepackage{a4wide}
\usepackage{colordvi,gnuplotcolors}

\newcommand{\gev}{\text{GeV}}
\newcommand{\tev}{\text{TeV}}
\newcommand{\fb}{\text{fb}}

\newcommand{\Sbot}{ {\tilde{b}} }

\newcommand{\mhpm}{ m_{H^{\pm}} }

\newcommand{\MSUSY}{ M_{\text{SUSY}}}

\newcommand{\tb}{\tan\beta}

\begin{document}

\title{\hfill {\normalsize hep-ph/0209124}\\
	\hfill {\normalsize MPI-PhT/2002-42}\\[1cm]
Associated $W^\pm H^\mp$ production at $e^+ e^-$ and hadron colliders
\thanks{Talk presented at the ''10th International Conference
on Supersymmetry and Unification of Fundamental Interactions, SUSY 02'',
DESY Hamburg, June 17-23 2002} }
\author{Oliver Brein\\
  {\em \small Max-Planck-Institut f\"ur Physik, M\"unchen}\\
  \small E-mail: obr@mppmu.mpg.de
  }

\date{\vspace*{.5cm}}

\maketitle

\begin{abstract}
We show some results 
in the minimal supersymmetric Standard Model (MSSM) 
and the two Higgs doublet model (THDM) 
for $W^\pm H^\mp$ production cross sections at $e^+ e^-$ and hadron colliders.
\footnote{The Fortran codes to calculate the cross section predictions 
         of the processes discussed here are available upon request.}
We demonstrate that the predictions for the cross sections 
in the two models can be vastly different.
Observing these processes, once the existence of a charged
Higgs is established, might shed some light on the underlying model for an
extended Higgs sector.
\end{abstract}

\section{Introduction}

\noindent {\bf Charged Higgs search and beyond:}\\
The discovery of a charged Higgs boson
at future colliders would be 
an unambiguous sign of an extended Higgs sector.
The production of charged Higgs bosons at the Tevatron and the LHC 
will happen mainly via the partonic
processes: $gb \to t H^+$ and $gg \to t b H^+$ \cite{plehnstalk}.
Other partonic processes like charged Higgs pair production
($gg/q\bar q \to H^+ H^-$) \cite{hadrohphm} 
and associated $W^\pm H^\mp$ production
($gg/b\bar b \to H^\pm W^\mp$) \cite{WH-hadro-first,ourgghw,WH-hadro-others} 
have also been considered, 
but they lead to a much smaller rate at these colliders.
At an $e^+ e^-$ collider the main production process
for charged Higgs bosons is pair production ($e^+ e^- \to H^+ H^-$)
which is mediated mainly via Photon- and $Z$- exchange in the
s-channel. 
If the collider energy is not sufficient for 
pair production 
(i.e. $\sqrt{s} < 2 \mhpm$) the associated $W^\pm H^\mp$ production
becomes dominant for $\sqrt{s} > m_W + \mhpm$. 
In this case 
the signal 
rate is much smaller because the process is effectively loop-induced.

The methods to discover the charged Higgs boson at the LHC 
seem to be well established by now \cite{latestHWGReport}.
Once it is found, one 
would like to 
learn more about the charged Higgs boson 
beyond its mere existence and its mass. 
But, the main production processes at LHC are rather insensitive  to
the underlying model. 
More interesting in this respect is the associated $W^\pm H^\mp$ production 
process.
At an $e^+ e^-$ collider the process is 
effectively loop-induced 
and at hadron colliders 
the gluon-fusion process
is loop-induced. Loop-induced processes
depend already at leading order 
on the virtual particles in the loops. 
Thus, with a cross section measurement of such a process one 
would gain information about the underlying model.

\smallskip

\label{param}
\noindent {\bf MSSM parameter scenario:}\\
We are interested in scenarios which potentially
show large effects from virtual squarks
and especially from Stops. Therefore, the MSSM parameter scenario used for 
the numerical results presented here has a rather low Squark mass scale
and shows large mixing in the Stop sector. 
We choose:
a) the common sfermion mass scale $\MSUSY = 350\,\gev$,
b) the Stop mixing parameter $X_t  = A_t - \mu \cot\beta = -800\,\gev$,
c) the supersymmetric Higgs mass parameter $\mu  = 300\,\gev$,
d)~the gaugino mass parameters $\approx {\cal O}(1\,\tev)$.

The main features of the resulting scenario are the following.
a) 
No mass bounds from direct search results are violated and
indirect constraints on the parameters coming from requiring vacuum stability 
and from experimental bounds on the electroweak 
precision observables are also
fulfilled. 
b) The mass of the light Stop is of order $100\,\gev$ and there is a
large mass splitting between the two Stop mass eigenstates.
c) The Stop mixing angle is approximately $45^\circ$. Thus maximal
mixing occurs, i.e. all entries in the Stop mixing matrix have approximately 
the same magnitude.
d) 
The mass of the lightest MSSM Higgs boson $m_h$ is
almost maximal with respect to $X_t$ 
(because $|X_t| \approx 2\MSUSY$ \cite{HHW}).
Note that this scenario allows for the production 
of Stop pairs at the energy scales considered in the following sections
and the decay
of the charged Higgs into Stop and Sbottom will be kinematically 
allowed for $\mhpm \gtrsim 450\,\gev$.

\section{$W^\pm H^\mp$ production at the LHC}

\begin{figure}[ht]
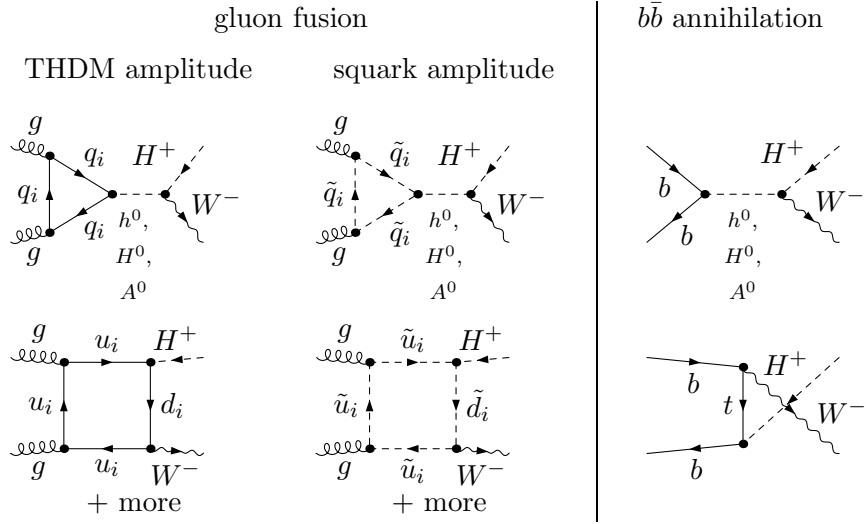

\begin{center}
\begin{tabular}[3]{cc|c}
\multicolumn{2}{c|}{{gluon fusion}} & {$b\bar b$ annihilation}\\[.2cm]
THDM amplitude & squark amplitude &\\
\begin{feynartspicture}(80,160)(1,2)
\FADiagram{}
\FAProp(0.,15.)(4.,14.)(0.,){/Cycles}{0}
\FALabel(2.54571,16.2028)[b]{$g$}
\FAProp(0.,5.)(4.,6.)(0.,){/Cycles}{0}
\FALabel(2.37593,4.47628)[t]{$g$}
\FAProp(20.,15.)(16.,10.)(0.,){/ScalarDash}{1}
\FALabel(17.2697,12.9883)[br]{$H^+$}
\FAProp(20.,5.)(16.,10.)(0.,){/Sine}{-1}
\FALabel(18.7303,7.98828)[bl]{$W^-$}
\FAProp(16.,10.)(10.5,10.)(0.,){/ScalarDash}{0}
\FALabel(13.25,9.18)[t]{$\begin{array}[1]{c}
	\scriptstyle h^0,\\
	\scriptstyle H^0,\\
	\scriptstyle A^0
			\end{array}$
			}
\FAProp(4.,14.)(4.,6.)(0.,){/Straight}{-1}
\FALabel(2.93,10.)[r]{$q_i$}
\FAProp(4.,14.)(10.5,10.)(0.,){/Straight}{1}
\FALabel(7.58235,12.8401)[bl]{$q_i$}
\FAProp(4.,6.)(10.5,10.)(0.,){/Straight}{-1}
\FALabel(7.58235,7.15993)[tl]{$q_i$}
\FAVert(4.,14.){0}
\FAVert(4.,6.){0}
\FAVert(16.,10.){0}
\FAVert(10.5,10.){0}

\FADiagram{}
\FAProp(0.,15.)(5.5,14.5)(0.,){/Cycles}{0}
\FALabel(2.95371,16.5108)[b]{$g$}
\FAProp(0.,5.)(5.5,5.5)(0.,){/Cycles}{0}
\FALabel(2.89033,4.18637)[t]{$g$}
\FAProp(20.,15.)(14.5,14.5)(0.,){/ScalarDash}{1}
\FALabel(17.1097,15.8136)[b]{$H^+$}
\FAProp(20.,5.)(14.5,5.5)(0.,){/Sine}{-1}
\FALabel(17.1097,4.18637)[t]{$W^-$}
\FAProp(5.5,14.5)(5.5,5.5)(0.,){/Straight}{-1}
\FALabel(4.43,10.)[r]{$u_i$}
\FAProp(5.5,14.5)(14.5,14.5)(0.,){/Straight}{1}
\FALabel(10.,15.57)[b]{$u_i$}
\FAProp(5.5,5.5)(14.5,5.5)(0.,){/Straight}{-1}
\FALabel(10.,4.43)[t]{$u_i$}
\FAProp(14.5,14.5)(14.5,5.5)(0.,){/Straight}{1}
\FALabel(15.57,10.)[l]{$d_i$}
\FAVert(5.5,14.5){0}
\FAVert(5.5,5.5){0}
\FAVert(14.5,14.5){0}
\FAVert(14.5,5.5){0}
\hspace*{1cm} + more
\end{feynartspicture}
\hspace{.7cm}

&

\begin{feynartspicture}(80,160)(1,2)
\FADiagram{}
\FAProp(0.,15.)(4.,14.)(0.,){/Cycles}{0}
\FALabel(2.54571,16.2028)[b]{$g$}
\FAProp(0.,5.)(4.,6.)(0.,){/Cycles}{0}
\FALabel(2.37593,4.47628)[t]{$g$}
\FAProp(20.,15.)(16.,10.)(0.,){/ScalarDash}{1}
\FALabel(17.2697,12.9883)[br]{$H^+$}
\FAProp(20.,5.)(16.,10.)(0.,){/Sine}{-1}
\FALabel(18.7303,7.98828)[bl]{$W^-$}
\FAProp(16.,10.)(10.5,10.)(0.,){/ScalarDash}{0}
\FALabel(13.25,9.18)[t]{$\begin{array}[1]{c}
	\scriptstyle h^0,\\
	\scriptstyle H^0,\\
	\scriptstyle A^0
			\end{array}$}
\FAProp(4.,14.)(4.,6.)(0.,){/ScalarDash}{-1}
\FALabel(2.93,10.)[r]{$\tilde q_i$}
\FAProp(4.,14.)(10.5,10.)(0.,){/ScalarDash}{1}
\FALabel(7.58235,12.8401)[bl]{$\tilde q_i$}
\FAProp(4.,6.)(10.5,10.)(0.,){/ScalarDash}{-1}
\FALabel(7.58235,7.15993)[tl]{$\tilde q_i$}
\FAVert(4.,14.){0}
\FAVert(4.,6.){0}
\FAVert(16.,10.){0}
\FAVert(10.5,10.){0}

\FADiagram{}
\FAProp(0.,15.)(5.5,14.5)(0.,){/Cycles}{0}
\FALabel(2.95371,16.5108)[b]{$g$}
\FAProp(0.,5.)(5.5,5.5)(0.,){/Cycles}{0}
\FALabel(2.89033,4.18637)[t]{$g$}
\FAProp(20.,15.)(14.5,14.5)(0.,){/ScalarDash}{1}
\FALabel(17.1097,15.8136)[b]{$H^+$}
\FAProp(20.,5.)(14.5,5.5)(0.,){/Sine}{-1}
\FALabel(17.1097,4.18637)[t]{$W^-$}
\FAProp(5.5,14.5)(5.5,5.5)(0.,){/ScalarDash}{-1}
\FALabel(4.43,10.)[r]{$\tilde u_i$}
\FAProp(5.5,14.5)(14.5,14.5)(0.,){/ScalarDash}{1}
\FALabel(10.,15.57)[b]{$\tilde u_i$}
\FAProp(5.5,5.5)(14.5,5.5)(0.,){/ScalarDash}{-1}
\FALabel(10.,4.43)[t]{$\tilde u_i$}
\FAProp(14.5,14.5)(14.5,5.5)(0.,){/ScalarDash}{1}
\FALabel(15.57,10.)[l]{$\tilde d_i$}
\FAVert(5.5,14.5){0}
\FAVert(5.5,5.5){0}
\FAVert(14.5,14.5){0}
\FAVert(14.5,5.5){0}
\hspace*{1cm} + more
\end{feynartspicture}
\hspace{.7cm} & 

\hspace{.2cm}
\begin{feynartspicture}(80,160)(1,2)
\FADiagram{}
\FAProp(0.,15.)(6.,10.)(0.,){/Straight}{1}
\FALabel(2.48771,11.7893)[tr]{$b$}
\FAProp(0.,5.)(6.,10.)(0.,){/Straight}{-1}
\FALabel(3.51229,6.78926)[tl]{$b$}
\FAProp(20.,15.)(14.,10.)(0.,){/ScalarDash}{1}
\FALabel(16.4877,13.2107)[br]{$H^+$}
\FAProp(20.,5.)(14.,10.)(0.,){/Sine}{-1}
\FALabel(17.5123,8.21074)[bl]{$W^-$}
\FAProp(6.,10.)(14.,10.)(0.,){/ScalarDash}{0}
\FALabel(10.,9.18)[t]{$\begin{array}[1]{c}
	\scriptstyle h^0,\\
	\scriptstyle H^0,\\
	\scriptstyle A^0
			\end{array}$}
\FAVert(6.,10.){0}
\FAVert(14.,10.){0}

\FADiagram{}
\FAProp(0.,15.)(10.,14.)(0.,){/Straight}{1}
\FALabel(4.84577,13.4377)[t]{$b$}
\FAProp(0.,5.)(10.,6.)(0.,){/Straight}{-1}
\FALabel(5.15423,4.43769)[t]{$b$}
\FAProp(20.,15.)(10.,6.)(0.,){/ScalarDash}{1}
\FALabel(16.8128,13.2058)[br]{$H^+$}
\FAProp(20.,5.)(10.,14.)(0.,){/Sine}{-1}
\FALabel(17.6872,8.20582)[bl]{$W^-$}
\FAProp(10.,14.)(10.,6.)(0.,){/Straight}{1}
\FALabel(9.03,10.)[r]{$t$}
\FAVert(10.,14.){0}
\FAVert(10.,6.){0}
\end{feynartspicture}
\end{tabular}\\[.3cm]
\end{center}
\caption{\label{LHC-HW-FD}
Typical Feynman diagrams for the partonic 
processes gluon fusion and $b \bar b$ annihilation contributing
to $W^\pm H^\mp$ production at a hadron collider.}
\end{figure}

There are two partonic processes contributing to $W^\pm H^\mp$
production at hadron colliders like the LHC
\footnote{The work presented in
this section has been done in collaboration 
with Wolfgang Hollik and Shinya Kanemura \cite{ourgghw}.}. 
On the one hand there 
is the tree-level process $b\bar b$ annihilation and on the 
other hand there is the loop-induced gluon fusion process
(see Figure \ref{LHC-HW-FD}). 
Both processes have been considered
already in \cite{WH-hadro-first}, but 
in the MSSM calculation squark loops had been neglected.
The full MSSM prediction for the gluon-fusion process 
appeared much later \cite{ourgghw,WH-hadro-others}.

$b\bar b$ annihilation has to be considered
if the bottom quark is 
treated as 
an active flavour in
the proton. 
In Figure \ref{LHC-HW-FD}
all Feynman diagrams contributing on tree-level to the amplitude 
are displayed. There are s-channel exchange diagrams for each of the neutral
Higgs bosons and a t-channel diagram including a virtual top quark.
Recently, the QCD corrections to the $b\bar b$ annihilation process
became known \cite{hollik-zhu}.

The gluon fusion process is mediated by loops of virtual quarks
and squarks (see Figure \ref{LHC-HW-FD}). Thus, the amplitude 
naturally splits into two subsets, one consisting of the diagrams 
with quark loops (THDM amplitude) and the other 
consisting of all diagrams with squark loops (squark amplitude). 
Figure \ref{LHC-HW-FD} shows essentially all
types of Feynman diagrams that contribute to the process in the MSSM.

The reason 
why we treat a tree-level and a loop-induced process
on equal footing 
is the much larger number of gluon fusion events compared to 
$b \bar b$ annihilation events at a high energy
proton collider.
Formally, the distributions of bottom quarks and anti-quarks 
are suppressed by one power of $\alpha_S $ compared to the 
gluon distribution which is of order one. This is because the bottom
is a heavy quark which doesn't have a valence distribution
in the proton. Therefore, its distribution is generated mainly by 
gluon splitting processes. By naively using the bottom quark 
distribution in order to estimate the hadronic cross section 
via $b \bar b$ annihilation it is well known that 
one overestimates the true cross section \cite{olness-tung}. The way to improve 
the prediction is to include additional processes of the same order in 
$\alpha_S$ like $gb \to bH^+W^-$ + c.c. or $gg \to b\bar b H^\pm W^\mp$.
But then a suitable subtraction of configurations with collinear 
bottom quarks of these additional processes has to be applied in order 
to avoid double-counting contributions which are already resummed in
the bottom quark distribution \cite{latestHWGReport,spira}. 
In our analysis we use for the $b \bar b$
process the LO cross section 
with a running bottom mass instead of the pole mass 
as an approximation to the NLO result \cite{hollik-zhu}
which leads to a considerable reduction of the hadronic cross section.
But still, this might be an overestimation.

\smallskip

\noindent {\bf MSSM results:}\\
Naturally, the $b \bar b$ cross section is rather insensitive to 
the superpartner spectrum as no SUSY contributions appear at LO.
The gluon fusion process shows, as expected, large sensitivity to 
the squark spectrum which appears in the loops. Especially,
the squark contribution to the amplitude can get large due to
threshold effects in the box loops.
A striking feature 
of the quark loop contribution to the gluon fusion amplitude, 
already noted in \cite{WH-hadro-first},
is a large destructive interference
between quark loops of box- and triangle-type. 
Both partonic processes, $b \bar b$ annihilation and gluon fusion, 
are rather sensitive to the neutral Higgs sector.
As a consequence, in the THDM the choice of the masses of the neutral Higgs
bosons can influence the cross section dramatically 
(see section \ref{THDM-LHC}), whereas in the MSSM the neutral Higgs
boson masses are essentially fixed, once values have been assigned
to $\mhpm$ and $\tb$.

\begin{figure}[hb]
  \psfrag{SIGMAPP}[c][c]{\huge $\sigma( pp \to H^\pm W^\mp) [\fb]$}
  \psfrag{TANB06}[l][l]{\huge $\tb = 6$}
  \psfrag{MHPM450}[l][l]{\huge $\mhpm = 450\,\gev$}
  \psfrag{MHPM}[c][b]{\huge $\mhpm [\gev]$}
  \psfrag{TANB}[c][b]{\huge $\tb$}
 \psfrag{BBBAR}[l][l]{\huge \Cyan{$ b\bar b$}}
 \psfrag{GGMSSM}[l][l]{\huge \Red{$ g g, \text{MSSM}$}}
 \psfrag{GG2HDM}[l][l]{\huge \Blue{$ g g, \text{sTHDM}$}}
  \psfrag{BBBAR2}[r][r]{\huge \Cyan{$ b\bar b$}}
 \psfrag{GGMSSM2}[l][l]{\huge \Red{$ g g, \text{MSSM}$}}
 \psfrag{GG2HDM2}[l][l]{\huge \Blue{$ g g, \text{sTHDM}$}}
 \resizebox*{.6\width}{.6\height}{\includegraphics*{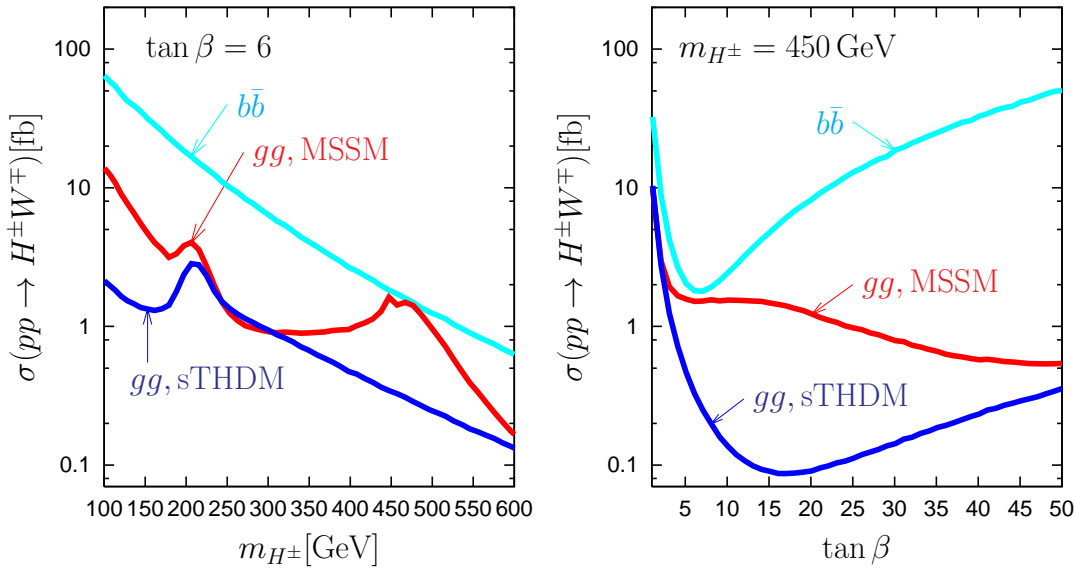}}
\caption{\label{LHC-HW-PLOT}
Hadronic cross section for $W^\pm H^\mp$ production via $b \bar b$
annihilation 
and gluon fusion 
as a function of $\mhpm$ and $\tb$.}
\end{figure}

Figure \ref{LHC-HW-PLOT} shows the MSSM prediction for the 
hadronic cross section for $W^\pm H^\mp$ production at the LHC.
The contributions from $b\bar b$ annihilation 
and gluon fusion are displayed separately as a function of the charged Higgs
mass $\mhpm$ and $\tb$. The results of the full MSSM (using the parameter
scenario described above) are compared to a MSSM-like THDM
(denoted by sTHDM in Figure \ref{LHC-HW-PLOT}). 
For the $b\bar b$
annihilation only
one line, valid for both cases, is shown.
The gluon
fusion process instead is significantly different in the full MSSM 
and the MSSM-like THDM. 
The cross section via gluon fusion in the full MSSM can be up to
an order of magnitude larger than in the sTHDM in our scenario. 
This is mainly due to threshold effects in the box-type squark loops 
which are not present in the sTHDM amplitude. 
The effect is maximal, if the production threshold 
($m_W+\mhpm$) is somewhat below the position in
$\sqrt{\hat s}$ ($= 2 m_{\Sbot_1}$ (here $\approx 680\,\gev$) )
where the threshold peak in the partonic cross section appears.
This happens in the left plot in Figure \ref{LHC-HW-PLOT} 
for $\mhpm \approx 450\,\gev$.
For this interesting value of $\mhpm$ 
the right plot in Figure \ref{LHC-HW-PLOT} shows the $\tb$-dependence.
One can see that the pronounced minimum in the MSSM-like THDM
has completely disappeared in the full MSSM and especially for 
$\tb$ up to about 15 the cross section for the gluon fusion process 
has the same order of magnitude as for the $b \bar b$ annihilation process.
Of course for large $\tb$ the $b \bar b$ annihilation process
dominates.
\smallskip

\label{THDM-LHC}
\noindent {\bf THDM results:}\\
In the MSSM the mass relations among the Higgs bosons for large $\mhpm$ lead
to 
\begin{align*}
m_{h^0} = {\cal O} ( 100 \gev),\; m_{H^0} \approx m_{A^0} \approx \mhpm \;.
\end{align*}
This has some interesting consequences for the process under study. 
a) The contributions of neutral Higgs s-channel exchange
cannot decouple from the process, because the masses of the 
heavy Higgs bosons are nearly degenerate.
b) Resonant s-channel Higgs exchange is impossible in the
MSSM, because the threshold $\mhpm+m_W$  is always larger than
any of the masses of the neutral Higgs bosons ($m_{h^0,H^0,A^0}$).
c) Therefore, the negative interference between box- and triangle-type
quark loops in the gluon fusion amplitude 
seems to be unavoidable in the MSSM.

But this is not true in the general THDM. As all Higgs masses are 
free parameters, there can be resonant s-channel exchange of neutral Higgs
bosons. It is also possible to choose some of the neutral Higgs bosons
much heavier than the threshold ($\mhpm+m_W$), thus suppressing
their contribution to the amplitude. 
Thus, in the 
general THDM the above-mentioned strong negative interference need
not be present and therefore the cross section can be much larger 
than in the MSSM.

\begin{figure}[hb]
\vspace*{-2.8cm}
\begin{center}
  \psfrag{MH03}[c][t]{\huge $m_{A^0} [\text{GeV}]$}
  \psfrag{MH01}[c][l][1][90]{\huge $m_{H^0} [\text{GeV}]$} 
  \psfrag{BBBAR}[l][c]{$b \bar{b}$ } 
  \psfrag{M0.4}[l]{\GNUPlotI{\bf\large 0.4}}
  \psfrag{M0.8}[l]{\GNUPlotH{\bf\large 0.8}}
  \psfrag{M10}[l]{\GNUPlotG{\bf\large 10}}
  \psfrag{M80}[l]{\GNUPlotF{\bf\large 80}}
  \psfrag{M100}[l]{\GNUPlotE{\bf\large 100}}
  \psfrag{M200}[l]{\GNUPlotD{\bf\large 200}}
  \psfrag{M400}[l]{\GNUPlotC{\bf\large 400}}
  \psfrag{M500}[l]{\GNUPlotB{\bf\large 500}}
\resizebox*{0.7\width}{0.7\height}
{\includegraphics*{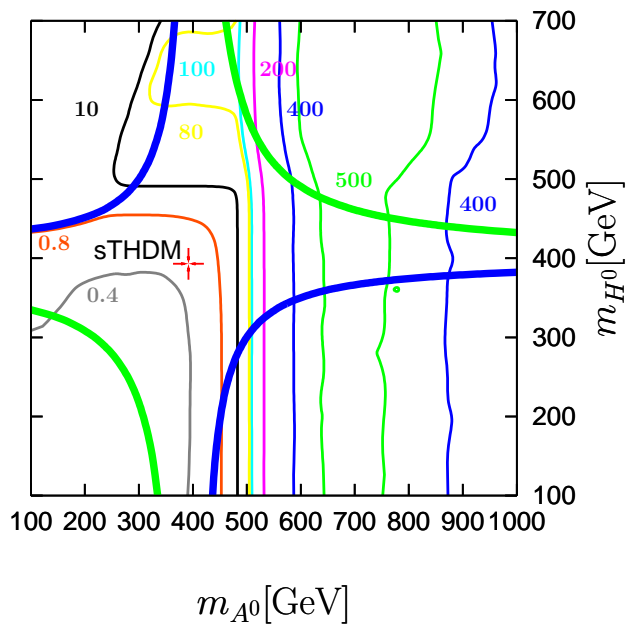}}
\end{center} 
\vspace*{-1.5cm}
\caption{\label{HW-THDM-PLOT}
Hadronic cross section for $W^\pm H^\mp$ production via gluon fusion
in $\fb$
as a function of the neutral Higgs masses $m_{A^0}$ and 
$m_{H^0}$ in the general THDM. The thin coloured lines represent
lines of equal cross section. The parameter regions between the thick lines
and
the corners 
of the plot are excluded (see text).
}
\end{figure}

In Figure \ref{HW-THDM-PLOT} one example for the typical 
behaviour in the general THDM is shown. We take  $\mhpm = 400\,\gev$,
$\tb = 6$ and the MSSM values 
for the mass of the lightest Higgs $m_{h^0}$ and
the mixing angle in the Higgs sector $\alpha$.
Figure \ref{HW-THDM-PLOT} shows a contour plot of the hadronic
cross section for $W^\pm H^\mp$ production via gluon fusion
as a function of 
the neutral Higgs masses $m_{A^0}$ and $m_{H^0}$. The point 
where the scenario coincides with a MSSM-like THDM
is marked by a red cross (labelled sTHDM). In this case the 
cross section via gluon fusion is $0.48\,\fb$ 
(via $b \bar b$ annihilation: $2.7\,\fb$).
The parameter region displayed in Figure \ref{HW-THDM-PLOT} is roughly
the region which is compatible with negative direct search 
results for $H^0$ and $A^0$ bosons and the requirement of tree-level unitarity.
Within the star-shaped region, bounded by the thick lines
in the plot, the scenarios are compatible with the measured value 
of the electroweak rho-parameter.
The cross section shows a steep rise if one of the Higgs masses,
$m_{A^0}$ or $m_{H^0}$, gets bigger than the $W^\pm H^\mp$
production threshold which is due to the resonant behavior of 
the corresponding s-channel exchange diagrams. But, even
without any resonance in the partonic cross section 
near
the $W^\pm H^\mp$ threshold
the cross section can be up to a factor 500 larger 
than in the corresponding MSSM-like THDM 
(e.g. for $m_{A^0}=800\,\gev$ and $m_{H^0}=400\,\gev$ we
have $\sigma = 470\,\fb$ via gluon fusion 
plus $860\,\fb$ via $b \bar b$ annihilation). 
It seems that some scenarios in the general THDM can be already
distinguished from the MSSM just by looking at the 
rate of $W^\pm H^\mp$ production.

\section{$e^+ e^- \to W^\pm H^\mp$}

\begin{figure}[ht]
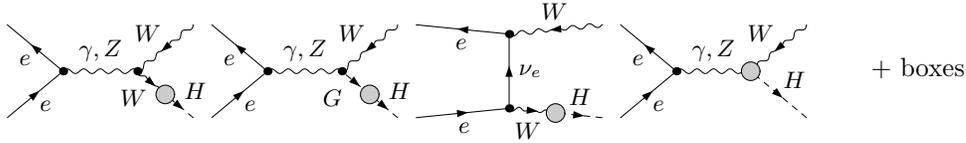

\begin{footnotesize}
\unitlength=1bp%
\begin{feynartspicture}(432,77)(5,1)
\FADiagram{}
\FAProp(0.,15.)(6.,10.)(0.,){/Straight}{-1}
\FALabel(2.48771,11.7893)[tr]{$e$}
\FAProp(0.,5.)(6.,10.)(0.,){/Straight}{1}
\FALabel(3.51229,6.78926)[tl]{$e$}
\FAProp(20.,15.)(14.,10.)(0.,){/Sine}{1}
\FALabel(16.4877,13.2107)[br]{$W$}
\FAProp(20.,5.)(17.,7.5)(0.,){/ScalarDash}{-1}
\FALabel(19.0123,6.96074)[bl]{$H$}
\FAProp(17.,7.5)(14.,10.)(0.,){/Sine}{-1}
\FALabel(14.9877,8.03926)[tr]{$W$}
\FAProp(6.,10.)(14.,10.)(0.,){/Sine}{0}
\FALabel(10.,11.07)[b]{$\gamma, Z$}
\FAVert(6.,10.){0}
\FAVert(14.,10.){0}
\FAVert(17.,7.5){-1}

\FADiagram{}
\FAProp(0.,15.)(6.,10.)(0.,){/Straight}{-1}
\FALabel(2.48771,11.7893)[tr]{$e$}
\FAProp(0.,5.)(6.,10.)(0.,){/Straight}{1}
\FALabel(3.51229,6.78926)[tl]{$e$}
\FAProp(20.,15.)(14.,10.)(0.,){/Sine}{1}
\FALabel(16.4877,13.2107)[br]{$W$}
\FAProp(20.,5.)(17.,7.5)(0.,){/ScalarDash}{-1}
\FALabel(19.0123,6.96074)[bl]{$H$}
\FAProp(17.,7.5)(14.,10.)(0.,){/ScalarDash}{-1}
\FALabel(14.,8.03926)[tr]{$G$}
\FAProp(6.,10.)(14.,10.)(0.,){/Sine}{0}
\FALabel(10.,11.07)[b]{$\gamma, Z$}
\FAVert(6.,10.){0}
\FAVert(14.,10.){0}
\FAVert(17.,7.5){-1}

\FADiagram{}
\FAProp(0.,15.)(10.,14.)(0.,){/Straight}{-1}
\FALabel(4.84577,13.4377)[t]{$e$}
\FAProp(0.,5.)(10.,6.)(0.,){/Straight}{1}
\FALabel(5.15423,4.43769)[t]{$e$}
\FAProp(20.,15.)(10.,14.)(0.,){/Sine}{1}
\FALabel(14.8458,15.5623)[b]{$W$}
\FAProp(20.,5.)(15.,5.5)(0.,){/ScalarDash}{-1}
\FALabel(17.6542,6.31231)[b]{$H$}
\FAProp(15.,5.5)(10.,6.)(0.,){/Sine}{-1}
\FALabel(12.,4.5)[t]{$W$}
\FAProp(10.,14.)(10.,6.)(0.,){/Straight}{-1}
\FALabel(11.07,10.)[l]{$\nu_e$}
\FAVert(10.,14.){0}
\FAVert(10.,6.){0}
\FAVert(15.,5.5){-1}

\FADiagram{}
\FAProp(0.,15.)(6.,10.)(0.,){/Straight}{-1}
\FALabel(2.48771,11.7893)[tr]{$e$}
\FAProp(0.,5.)(6.,10.)(0.,){/Straight}{1}
\FALabel(3.51229,6.78926)[tl]{$e$}
\FAProp(20.,15.)(14.,10.)(0.,){/Sine}{1}
\FALabel(16.4877,13.2107)[br]{$W$}
\FAProp(20.,5.)(14.,10.)(0.,){/ScalarDash}{-1}
\FALabel(17.5123,8.21074)[bl]{$H$}
\FAProp(6.,10.)(14.,10.)(0.,){/Sine}{0}
\FALabel(10.,11.07)[b]{$\gamma, Z$}
\FAVert(6.,10.){0}
\FAVert(14.,10.){-1}

\FADiagram{\vspace*{1.2cm}+ boxes}
\end{feynartspicture}
\end{footnotesize}

\caption{\label{EE-HW-FD}
Feynman diagrams contributing to the amplitude
of the process 
$e^+ e^- \to W^+ H^-$. Only generic one-loop vertex insertions are
shown.
}
\end{figure}

The leading contributions to the amplitude are one-loop Feynman diagrams
(see Figure \ref{EE-HW-FD})
\footnote{The work presented in this section has been done
in collaboration with Thomas Hahn and Wolfgang Hollik \cite{oureehw}.}.
There are diagrams with $H$--$W$ and $G$--$H$ mixing self energy insertions and
there are triangle- and box-type diagrams which contain either 
only THDM particles or superpartners in the loop.
The calculation of the cross section in the framework of the THDM
has been done by several authors \cite{eehw-thdm} and only 
recently the MSSM calculation has been done \cite{logan-su,oureehw}.

\begin{figure}[t]
\begin{center}
  \psfrag{SIGMA}[c][c]{\huge $\sigma( e^+ e^-\to H^\pm W^\mp) [\fb]$}
  \psfrag{TANB30}[l][l]{\huge $ \tb = 30$}
  \psfrag{MHPM350}[l][l]{\huge $ \mhpm = 350\,\gev$}
  \psfrag{SQRTS0500}[l][l]{\huge $ \sqrt{s}\; = 500\,\gev$}
  \psfrag{MHPM}[c][b]{\huge $ \mhpm [\gev]$}
  \psfrag{TANB}[c][b]{\huge $ \tb$}
 \psfrag{MSSM}[l][l]{\Red{\huge $ \text{MSSM}$}}
 \psfrag{2HDM}[l][l]{\Blue{\huge $ \text{sTHDM}$}}
 \psfrag{GGMSSM2}[l][l]{\Red{\huge $ g g, \text{MSSM}$}}
 \psfrag{GG2HDM2}[l][l]{\Blue{\huge $ g g, \text{sTHDM}$}}
 \resizebox*{.6\width}{.6\height}{\includegraphics*{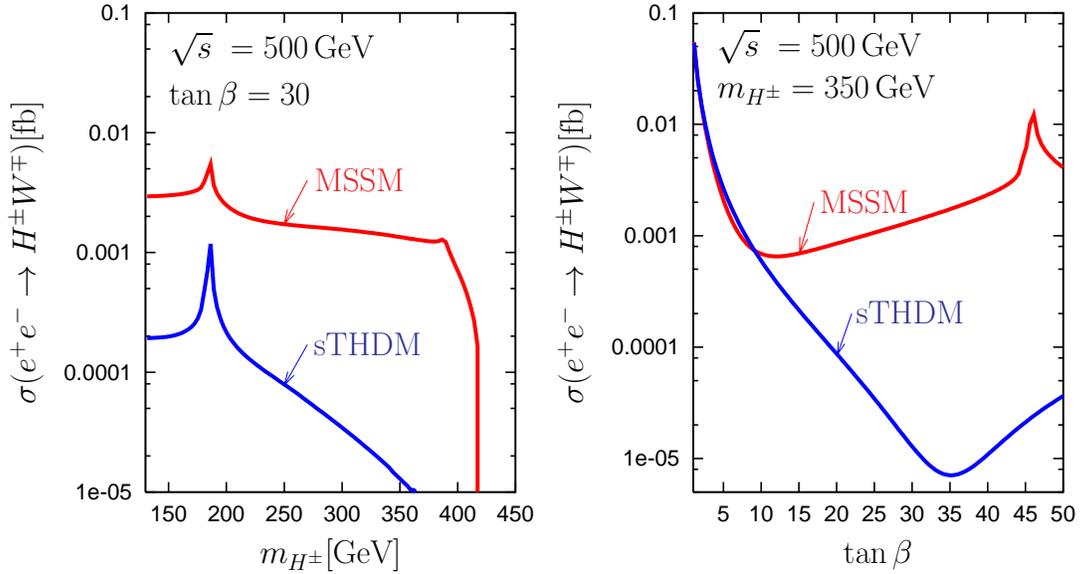}}
\end{center}
\caption{\label{EE-HW-PLOT}
Unpolarized cross section for $e^+ e^-\to H^\pm W^\mp$ for a 
center of mass energy of $\sqrt{s} = 500\,\gev$ as 
function of $\mhpm$ and $\tb$. The predictions
of the MSSM in our scenario 
and the corresponding MSSM-like 
THDM (sTHDM)
are compared.
}
\end{figure}

\noindent {\bf MSSM results:}\\
Typically the vertex diagrams give the main contribution to the cross 
section. Especially, vertex graphs with loops
of third generation quarks and squarks which couple to the 
outgoing charged Higgs boson are dominant.
In Figure \ref{EE-HW-PLOT} the predicted cross section 
for the process $e^+ e^-\to W^\mp H^\pm$ is displayed 
for the full MSSM using the scenario of section \ref{param}
and the corresponding MSSM-like 
THDM (sTHDM). 
The Figure shows examples for the 
$\mhpm$- and $\tb$-dependence of the cross section.
Generically, for large $\tb$
the result for our MSSM scenario
is considerably larger than for the corresponding 
THDM (up to two orders of magnitude), as exemplified by the 
$\tb$-plot in Figure \ref{EE-HW-PLOT}.
The enhancement with respect to the sTHDM case is especially 
due to the squark-loop diagrams which contain enhanced couplings 
to the charged Higgs.
The $\mhpm$-plot in Figure \ref{EE-HW-PLOT} shows the cross section for 
$\tb = 30$, i.e. in the area of large enhancement with respect to
$\tb$. It is interesting that the MSSM prediction stays at least one order of
magnitude larger than the sTHDM prediction 
over the entire charged Higgs mass range which is kinematically accessible.

The large differences between the MSSM prediction in our scenario
and the corresponding THDM prediction
suggest that, once a charged Higgs is found, one might 
gain valuable information about the underlying model it is embedded in
by observing the associated production channel.
The results presented here assume unpolarized beams. Note that
optimal polarization of $e^-$ and $e^+$ can increase the cross section
almost by a factor of four. It seems that large corrections by 
neutralino and chargino loops are possible, 
but 
this issue is still under investigation \cite{oureehw}.

\section{Summary}

We investigated the predictions of the MSSM and the THDM for $W^\pm H^\mp$
production at the LHC and a future 
high energy 
$e^+ e^-$
collider with $500\,\gev$
center of mass energy. 
One interesting
result for the hadron collider is that among the contributing partonic
processes the loop-induced gluon-fusion can compete with 
the tree-level $b\bar b$
annihilation in certain cases. Therefore and with regard to the potential
pitfall of overestimating the $b\bar b$ cross section, a careful treatment
of the $b\bar b$ process is needed (see \cite{hollik-zhu,spira}).
Furthermore, we showed that the hadronic cross section in the general THDM
can be much larger than in the MSSM (about a factor 500-1000). 
At a high energy $e^+ e^-$ collider the $W^\pm H^\mp$ 
production is especially important for the charged Higgs boson
detection, if pair production is kinematically forbidden.

In general these loop-induced processes are highly model-dependent.
Therefore, they are well suited to gain information about 
the underlying model.
The MSSM scenarios with large Stop mixing and low sfermion mass scale,
for which we showed here one example, can give rise to cross sections 
for $W^\pm H^\mp$ production at the LHC or $e^+ e^-$ machines 
which are significantly different from a MSSM-like THDM.

\bigskip

\section*{Acknowledgement}
The results discussed here were obtained in collaboration with
Shinya Kanemura, Wolfgang Hollik and Thomas Hahn.
I thank the organizers of SUSY 2002 for creating a 
very pleasant meeting.

\end{document}